\documentclass[lettersize,journal]{IEEEtran}
\usepackage{amsmath,amsfonts}
\usepackage{algorithmic}
\usepackage{algorithm}
\usepackage{array}
\usepackage[caption=false,font=normalsize,labelfont=sf,textfont=sf]{subfig} 
\usepackage{textcomp}
\usepackage{stfloats}
\usepackage{url}
\usepackage{verbatim}
\usepackage{graphicx}  
\usepackage{float}      

\usepackage{booktabs}
\usepackage{url}
\usepackage[numbers]{natbib}
\usepackage{amssymb}

\begin{document}

\title{Multimodal Emotion Recognition using Audio-Video Transformer Fusion with Cross Attention}

\author{Shravan Venkatraman, Joe Dhanith P R, Vigya Sharma, Santhosh Malarvannan

\thanks{This work did not receive any funding. (Corresponding author: Joe Dhanith P R.)}
\thanks{Shravan Venkatraman is with the School of Computer Science and Engineering, Vellore Institute of Technology (VIT) University, Chennai, India (e-mail: shravan.venkatraman2021@vitstudent.ac.in).}
\thanks{Joe Dhanith P R is with the School of Computer Science and Engineering, Vellore Institute of Technology (VIT) University, Chennai, India (e-mail: joedhanith.pr@vit.ac.in).}
\thanks{Vigya Sharma is with the School of Computer Science and Engineering, Vellore Institute of Technology (VIT) University, Chennai, India (e-mail: vigyas216@gmail.com).}
\thanks{Santhosh Malarvannan is with the School of Computer Science and Engineering, Vellore Institute of Technology (VIT) University, Chennai, India (e-mail: santhosh.malarvannan2021@vitstudent.ac.in).}
}



\maketitle

\begin{abstract}
Multimodal emotion recognition (MER) aims to infer human affect by jointly modeling audio and visual cues; however, existing approaches often struggle with temporal misalignment, weakly discriminative feature representations, and suboptimal fusion of heterogeneous modalities. To address these challenges, we propose \textit{AVT-CA}, an Audio--Video Transformer architecture with Cross Attention for robust emotion recognition. The proposed model introduces a hierarchical video feature representation that combines channel attention, spatial attention, and local feature extraction to emphasize emotionally salient regions while suppressing irrelevant information. These refined visual features are integrated with audio representations through an intermediate transformer-based fusion mechanism that captures interlinked temporal dependencies across modalities. Furthermore, a cross-attention module selectively reinforces mutually consistent audio--visual cues, enabling effective feature selection and noise-aware fusion. Extensive experiments on three benchmark datasets—CMU-MOSEI, RAVDESS, and CREMA-D—demonstrate that AVT-CA consistently outperforms state-of-the-art baselines, achieving significant improvements in both accuracy and F1-score. Our source code is publicly available at \url{https://github.com/shravan-18/AVTCA}.
\end{abstract}


\section{Introduction}
\IEEEPARstart{E}{ffective} human communication hinges on the fundamental skill of emotional awareness, enabling individuals to comprehend and respond to the feelings of others. Conventional emotion recognition methods \cite{1} often rely on data from a single source, such as speech patterns or facial expressions. However, human emotional communication is inherently multimodal, involving a complex interplay of verbal and non-verbal cues. To address this complexity, recent research has increasingly focused on MER tasks that combine audio and video sources. By leveraging both verbal intonations and facial expressions \cite{2}, these approaches aim to provide a deeper and more comprehensive understanding of an individual's emotional state.

Consider the example: "I'm really happy for you!" Determining whether genuine happiness is being expressed based solely on the tone of voice can be challenging. Visual cues, such as a sincere smile, wrinkled eyes, and an overall positive demeanor, offer additional insights. However, relying solely on visual data has its drawbacks, as a smile can be faked to mask true feelings. Therefore, combining audio and visual elements is essential to capture a more accurate emotional context \cite{3}, \cite{4}.

\begin{figure*}[htbp]  
    \centering
    \subfloat[Angry]{%
        \includegraphics[width=0.3\textwidth,height=4cm,keepaspectratio=false]{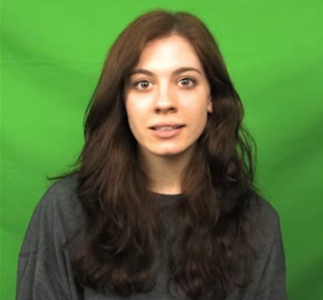}%
        \label{fig:image1}
    }
    \hfill
    \subfloat[Disgust]{%
        \includegraphics[width=0.3\textwidth,height=4cm,keepaspectratio=false]{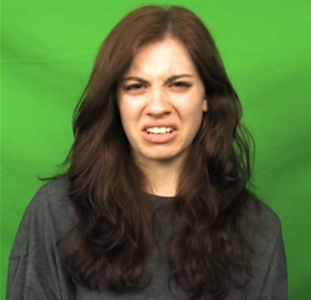}%
        \label{fig:image2}
    }
    \hfill
    \subfloat[Fear]{%
        \includegraphics[width=0.3\textwidth,height=4cm,keepaspectratio=false]{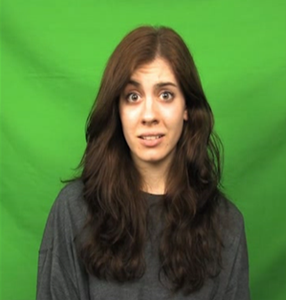}%
        \label{fig:image3}
    }

    \subfloat[Happy]{%
        \includegraphics[width=0.3\textwidth,height=4cm,keepaspectratio=false]{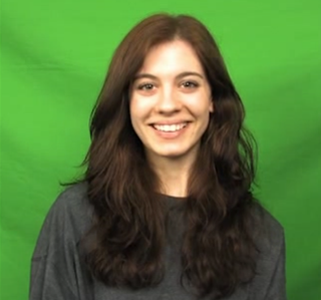}%
        \label{fig:image4}
    }
    \hfill
    \subfloat[Neutral]{%
        \includegraphics[width=0.3\textwidth,height=4cm,keepaspectratio=false]{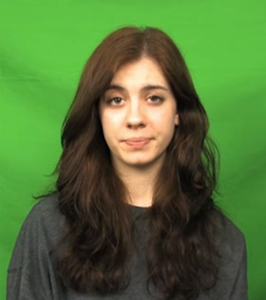}%
        \label{fig:image5}
    }
    \hfill
    \subfloat[Sad]{%
        \includegraphics[width=0.3\textwidth,height=4cm,keepaspectratio=false]{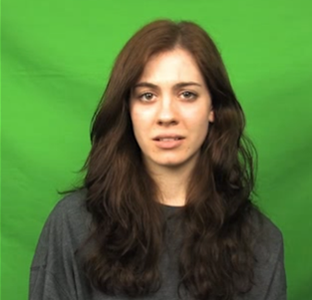}%
        \label{fig:image6}
    }

    \caption{
Examples of facial expressions from the CMU-MOSEI dataset.
}
    \label{fig:all_images}
\end{figure*}

Accurately recognizing emotions from multiple sources requires effective integration of audio and video information. This task is challenging due to three primary concerns: (i) synchronization, (ii) feature extraction, and (iii) fusion. Synchronizing audio and video data is crucial but technically demanding, especially when dealing with temporal inconsistencies and differing sample rates. Combining data from these sources in a way that maximizes their respective strengths is also challenging. Existing deep learning models struggle to determine the appropriate weight for each modality during fusion, and they must account for individual variations in emotional expression to avoid bias, necessitating careful feature extraction.

Early attempts at MER, which either integrated features from each source (early fusion) or combined final predictions (late fusion), yielded limited success. Newer techniques leveraging deep learning, particularly those involving attention mechanisms and transformers, have shown promise in exploring more sophisticated fusion methods. Extracting necessary features in unrestricted environments remains difficult, introducing uncertainty into the processing pipeline. This is especially problematic for language-based approaches, such as text transcriptions of audio signals, which are rarely available in real-world scenarios and require separate estimation. Pre-extracted features \cite{5}, \cite{6}, \cite{7}, commonly used in transformer-based designs, have become popular for MER, as they eliminate the need for end-to-end feature learning \cite{8}.

To address these challenges, this work proposes AVT-CA, a novel method for effective MER. AVT-CA introduces a unique feature extraction mechanism for video sources, comprising (i) channel attention, (ii) spatial attention, and (iii) a local feature extractor. Channel attention assesses the importance of each feature channel in the input video, spatial attention identifies critical regions for emotion recognition, and the local feature extractor refines the frames selected by both channel and spatial attention for efficient feature representation. This approach eliminates the need for end-to-end feature extraction.

By utilizing the proposed feature representation module, the AVT-CA model can learn complex relationships between features through an intermediate-level fusion technique called Transformer Fusion. This approach addresses synchronization issues by capturing interlinked features between audio and video sources. The cross-attention mechanism in AVT-CA extracts significant features while discarding insignificant ones from both audio and video inputs, thereby resolving feature extraction and fusion challenges in MER tasks.

The key contributions of the proposed work are as follows:
\begin{itemize}
    \item AVT-CA introduces an innovative approach to audio-video emotion recognition by employing a novel feature representation strategy before learning from raw facial and speech data. This strategy encompasses channel attention, spatial attention, and a local feature extractor, which collectively enhance the model’s ability to discern intrinsic correlations within the data while mitigating the impact of preprocessing inaccuracies.

    \item AVT-CA presents a novel intermediate transformer fusion method that integrates inputs from complementary modalities post-initial feature extraction. This integration facilitates early learning of significant features from both modalities. Incorporating transformer blocks within each branch captures complex interrelations in the data, yielding robust and discriminative feature representations for precise emotion recognition.
    
    \item The proposed model introduces a unique cross attention mechanism after the intermediate transformer fusion. This method selects the most significant features within each modality based on their agreement with one another, prioritising features that reinforce each other during prediction. This agreement-driven fusion produces better discriminative representations, improving emotion recognition ability.

\end{itemize}

The rest of the paper is organized as follows: Section II reviews the literature in MER tasks, Section III introduces the proposed AVT-CA model, Section IV discusses the experimental setup, Section V analyzes the results and Section 6 concludes the paper.

\section{Related Work}

\subsection{Multimodal Learning for Emotion Recognition}
Emotion recognition has been an active area in machine learning, with growing interest in leveraging multimodal signals to improve performance. Foundational studies explore \emph{multimodal and multi-task learning}~\cite{9}, \emph{cross-modal generation}~\cite{10,11}, \emph{vision--language models}~\cite{12,13}, \emph{vision--audio integration}~\cite{14,15,16,17,18}, and \emph{zero-shot transfer}~\cite{19,20}. Multi-task approaches exploit complementary cues across data types, cross-modal generation maps between modalities, and vision–language/audio models support joint reasoning over heterogeneous inputs. These advances underpin efforts to build systems that integrate visual, acoustic, and textual information for richer affect analysis.

\subsection{Learning from Video with Weak or Noisy Supervision}
Videos contain diverse cues such as visual dynamics, speech, and context; yet large-scale annotation is expensive. Miech \emph{et al.}~\cite{21} aligned video and text representations using contrastive learning, releasing a 100M narrated-video dataset with ASR transcripts and subtitles. Amrani \emph{et al.}~\cite{22} modeled uncertainty through multimodal density estimation; Miech \emph{et al.}~\cite{23} proposed \emph{MIL-NCE}, and Alwassel \emph{et al.}~\cite{24} exploited unsupervised clustering. Beyond bimodal settings, several works~\cite{25,26,27,28,29} learn unified representations across vision, audio, and text, including \emph{Multi-Modal Versatile Networks}~\cite{26}, a shared embedding by Rouditchenko \emph{et al.}~\cite{29}, and clustering-enhanced objectives by Chen \emph{et al.}~\cite{27}. These methods highlight how contrastive and unsupervised strategies enable large-scale multimodal learning despite label noise.

\subsection{Emotion Representations and Transformer-based Fusion}
Recent research increasingly focuses on how to represent and fuse emotions across modalities. Dai \emph{et al.}~\cite{30} transferred textual emotion embeddings to visual and acoustic domains, while Hazarika \emph{et al.} introduced \emph{MISA} to disentangle modality-invariant and modality-specific cues~\cite{6}. Han \emph{et al.} developed a text-centered fusion transformer~\cite{5}, and Vu \emph{et al.}~\cite{31} addressed sparse labels via knowledge-distillation for expression and valence–arousal estimation. Other approaches leverage mutual-information maximization~\cite{7}, curated hand-crafted feature sets used by downstream techniques~\cite{32}, autoencoders with LSTMs for temporal affect~\cite{33}, and dataset restructuring for end-to-end MER from raw signals~\cite{34}.

Transformer and attention mechanisms have been extensively impactful in representation learning and multimodal fusion \cite{transformer1,transformer2}. Parthasarathy \emph{et al.}~\cite{35} modeled speech–face interactions with a multimodal transformer; Hazarika \emph{et al.} dynamically weighted cues~\cite{36}; and further variants include co-attention for audio–visual links~\cite{25}, cross-modal encoders for video–text~\cite{37,38}, and temporal–spatial integration~\cite{39}. Bottleneck transformers~\cite{40}, self-supervised training~\cite{41}, and facial-specific models such as the Distract-Your-Attention Network~\cite{42} or vector-quantized autoencoding~\cite{43} refine attention on subtle expressions. Unified backbones spanning video, audio, and text~\cite{44,45}, tensor-based fusion~\cite{46,47}, temporal alignment with experts~\cite{45,48}, CLIP-based retrieval transformers~\cite{49}, and residual self-attention networks for MER~\cite{50} further extend this line.

The review of prior work highlights three main limitations:
\begin{itemize}
    \item \textbf{Temporal misalignment:} Many systems suffer when audio and video streams are not perfectly synchronized, causing them to overlook short-lived or phase-shifted cues essential for accurate emotion recognition.
    \item \textbf{Weakly discriminative features:} Several approaches fail to capture fine-grained emotional cues, especially for subtle or compound emotions, because their feature representations are not sufficiently expressive.
    \item \textbf{Ineffective fusion:} Simple concatenation or early-fusion strategies often cannot model the complex dependencies between modalities, leading to suboptimal integration of visual and acoustic information.
\end{itemize}

\section{Proposed Work: \textit{AVTCA}}
This section details our proposed Audio-Video Transformer Fusion with Cross Attention (AVT-CA) model. The model employs a dual-channel approach $\mathcal{F}: \mathcal{X}_A \times \mathcal{X}_V \rightarrow \mathcal{Y}$ to process audio and visual data independently, leveraging complementary emotional cues in both modalities. Each channel extracts salient features $\mathbf{f}_A \in \mathbb{R}^{d_A}$ and $\mathbf{f}_V \in \mathbb{R}^{d_V}$ from its respective input stream.

The audio channel captures spectral-temporal characteristics including pitch variations $\Delta p(t)$, intensity $I(t)$, and rhythm $r(t) = \frac{1}{T}\sum_{i=1}^{T}|s(i) - s(i-1)|$. The visual channel processes facial expressions $\mathbf{E}_f \in \mathbb{R}^{n \times k}$, micro-expressions $\mathbf{M}_e$ with $\delta t < 500$ ms, and bodily gestures $\mathbf{G}(t)$. By processing audio $\mathbf{A} \in \mathbb{R}^{T \times C}$ and visual inputs $\mathbf{V} \in \mathbb{R}^{T \times H \times W \times 3}$ separately before fusion, our model preserves modality-specific temporal dynamics $\tau_m(t)$ and feature representations through transformations $\phi_A: \mathbb{R}^{T \times C} \rightarrow \mathbb{R}^{d_A}$ and $\phi_V: \mathbb{R}^{T \times H \times W \times 3} \rightarrow \mathbb{R}^{d_V}$.

\begin{figure*}[htbp]
\centering
\includegraphics[width=0.7\textwidth]{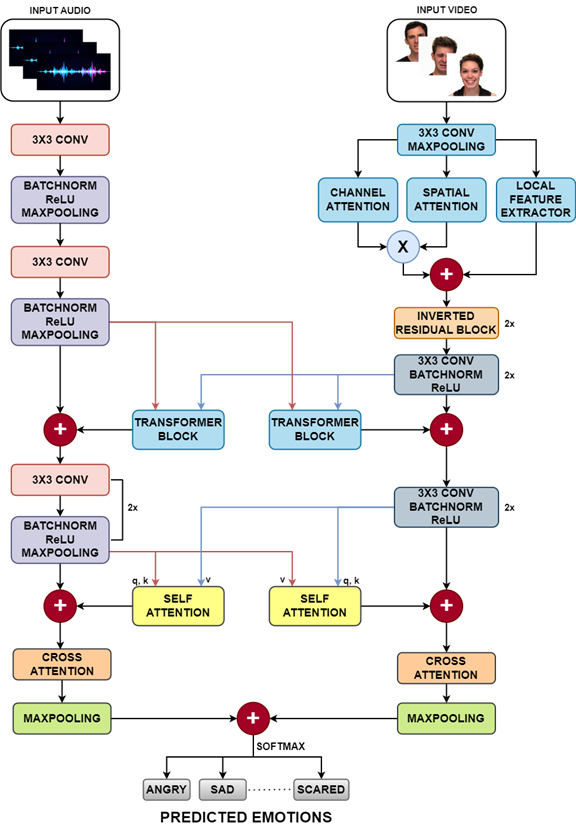}
\caption{
Architecture of the proposed \textbf{AVT-CA} model.
Audio and video inputs are processed by modality-specific convolutional blocks, refined through attention mechanisms, and fused via intermediate transformers and cross-attention to produce final emotion predictions.
}
\label{fig:avtca}
\end{figure*}

The overall architecture is illustrated in Figure \ref{fig:avtca}, with parallel processing streams integrated via our fusion mechanism $\mathcal{F}_{fusion}(\mathbf{f}_A, \mathbf{f}_V) = \sigma(\mathbf{W}[\mathbf{f}_A \oplus \mathbf{f}_V] + \mathbf{b})$. Its algorithmic workflow is presented in Algorithm~\ref{alg:AVTCA-alg}.

\subsection{Audio and Video Feature Extraction}

\subsubsection{Audio Processing Pipeline}

The input audio signal $\mathbf{A} \in \mathbb{R}^{T \times C}$ undergoes two consecutive convolutional blocks $\mathcal{B}_{a1}$ and $\mathcal{B}_{a2}$ that extract hierarchical features combining local $\mathbf{f}_{a,l}^{loc} \in \mathbb{R}^{d_{loc}}$ and global patterns $\mathbf{f}_{a,l}^{glob} \in \mathbb{R}^{d_{glob}}$ where $\mathbf{f}_{a,l} = [\mathbf{f}_{a,l}^{loc} \oplus \mathbf{f}_{a,l}^{glob}]$.

In the first block, 1D convolution is applied: $\mathbf{A}' = \text{Conv1D}(\mathbf{A}, \mathbf{W}_a, \mathbf{b}_a) = \sum_{j=0}^{k-1} \mathbf{A}_{i+j} \cdot \mathbf{W}_{a,j} + \mathbf{b}_a$, where $\mathbf{W}_a \in \mathbb{R}^{k \times C \times C'}$ and $\mathbf{b}_a \in \mathbb{R}^{C'}$ for kernel size $k=3$. This identifies temporal patterns such as pitch changes $\Delta p(t) = p(t) - p(t-\delta)$, timbre $\mathcal{T}(\omega) = \sum_{i=1}^{n} a_i \sin(2\pi \omega_i t + \phi_i)$, and intensity $I(t) = |A(t)|^2$.

Batch normalization stabilizes learning: $\mathbf{\hat{A}} = \gamma \frac{\mathbf{A}' - \mu_\mathcal{B}}{\sqrt{\sigma_\mathcal{B}^2 + \epsilon}} + \beta$. ReLU activation introduces non-linearity: $\mathbf{\tilde{A}} = \max(0, \mathbf{\hat{A}})$. Max pooling retains salient features: $\mathbf{\bar{A}} = \max_{i \in \{1,\dots,p\}} (\mathbf{\tilde{A}}_{[i:i+p-1]})$, focusing on dominant patterns $\mathcal{P}_d = \{\mathbf{p} \in \mathcal{P} | \mathcal{S}(\mathbf{p}) > \tau\}$.

The output $\mathbf{\bar{A}} \in \mathbb{R}^{\lfloor\frac{T}{p}\rfloor \times C'}$ enters a second identical block:
\begin{subequations}
    \begin{align}
        & \mathbf{A}_2' = \sum_{j=0}^{k-1} \mathbf{\bar{A}}_{i+j} \cdot \mathbf{W}_{a,j}^2 + \mathbf{b}_a^2, \\
        & \mathbf{\hat{A}}_2 = \gamma_2 \frac{\mathbf{A}_2' - \mu_{\mathcal{B},2}}{\sqrt{\sigma_{\mathcal{B},2}^2 + \epsilon}} + \beta_2, \\
        & \mathbf{\tilde{A}}_2 = \max(0, \mathbf{\hat{A}}_2), \\
        & \mathbf{\bar{A}}_2 = \max_{i \in \{1,\dots,p\}} (\mathbf{\tilde{A}}_{2,[i:i+p-1]}).
    \end{align}
\end{subequations}

This two-stage process $\mathcal{H}(\mathbf{A}) = \mathcal{B}_{a2}(\mathcal{B}_{a1}(\mathbf{A}))$ refines emotional state recognition, capturing low-level features (frequency variations $\nabla_t \omega(t)$, energy dynamics $E(t)$) and higher-level abstractions (prosody $\mathcal{P}(t)$, rhythm). These features distinguish emotions like anger ($\max_t |\nabla_t^2 p(t)| > \tau_a$), sadness ($\{r(t) < \tau_r, E(t) < \tau_E\}$), and fear ($\sigma^2_{\delta t} > \tau_f$). The output $\mathbf{\bar{A}}_2 \in \mathbb{R}^{\lfloor\frac{\lfloor\frac{T}{p}\rfloor}{p}\rfloor \times C''}$ feeds into transformer blocks for cross-modal interaction through attention mechanism $\mathcal{A}(\mathbf{Q}, \mathbf{K}, \mathbf{V}) = \text{softmax}(\frac{\mathbf{Q}\mathbf{K}^T}{\sqrt{d_k}})\mathbf{V}$.

\subsubsection{Video Processing Pipeline}

The video input $\mathbf{V} \in \mathbb{R}^{H \times W \times C}$ undergoes processing to extract spatial features, emphasize critical regions $\mathcal{R}_c = \{r_i \in \mathcal{R} | \mathcal{I}(r_i) > \tau_r\}$, and refine local details through transformation $\phi_V: \mathbb{R}^{H \times W \times C} \rightarrow \mathbb{R}^{d_V}$.

\paragraph{Initial Feature Extraction}
The input video tensor passes through convolution: $\mathbf{V}' = \text{Conv2D}(\mathbf{V}, \mathbf{W}_v, \mathbf{b}_v) = \sum_{m=0}^{k-1}\sum_{n=0}^{k-1} \mathbf{V}_{i+m,j+n} \cdot \mathbf{W}_{v,m,n} + \mathbf{b}_v$, capturing spatial features such as edges $\mathbf{E}(i,j) = \sqrt{G_x^2(i,j) + G_y^2(i,j)}$ and textures. Max pooling then reduces dimensions: $\mathbf{\bar{V}} = \max_{(p,q) \in \mathcal{W}_{i,j}} (\mathbf{V}'(p,q))$.

\paragraph{Attention Mechanisms}
Channel attention recalibrates feature importance: $\mathbf{\tilde{V}}_c = \sigma(\mathbf{W}_{fc2}\sigma(\mathbf{W}_{fc1}\frac{1}{H'W'}\sum_{i,j}\mathbf{\bar{V}}(i,j) + \mathbf{b}_{fc1}) + \mathbf{b}_{fc2})$, emphasizing channels with importance weighting $\omega_c = \frac{e^{\mathcal{S}(c)}}{\sum_{k} e^{\mathcal{S}(k)}}$. Spatial attention generates an attention map: $\mathbf{\tilde{V}}_s = \frac{e^{\mathbf{Z}(i,j)}}{\sum_{p,q} e^{\mathbf{Z}(p,q)}}$ where $\mathbf{Z} = \text{Conv2D}(\mathbf{\bar{V}}, \mathbf{W}_s, \mathbf{b}_s)$, assigning higher weights to emotionally significant regions. These are combined: $\mathbf{\tilde{V}} = \mathbf{\tilde{V}}_c \odot \mathbf{\tilde{V}}_s = \{\mathbf{\tilde{V}}_c(k) \cdot \mathbf{\tilde{V}}_s(i,j,1) \cdot \mathbf{\bar{V}}(i,j,k)\}$, integrating global and local feature importance.

\paragraph{Local Feature Extraction}
A local feature extractor divides frames into $N$ patches $\mathcal{P} = \{P_1, P_2, ..., P_N\}$ where $P_i \in \mathbb{R}^{\frac{H'}{n} \times \frac{W'}{m} \times C'}$: $\textbf{o}_i = \sum_{p,q} P_i(u+p,v+q) \cdot \mathbf{W}_{p}(p,q) + \mathbf{b}_{p}$. These features are concatenated: $\textbf{o}_{local} = [\textbf{o}_1, \textbf{o}_2, ..., \textbf{o}_N] \in \mathbb{R}^{N \times d_p}$ and combined with attention outputs: $\mathbf{\hat{V}} = \mathbf{\tilde{V}} + \psi(\textbf{o}_{local})$.

\paragraph{Inverted Residual Block}
The enhanced representation $\mathbf{\hat{V}}$ passes through two consecutive inverted residual blocks $\mathcal{B}_{inv}^1 \circ \mathcal{B}_{inv}^2$ using depthwise separable convolutions. This reduces parameters from $C' \times C' \times k \times k$ to $C' \times k \times k + C' \times C'$ while ensuring efficient gradient flow $\nabla_{\mathbf{x}} \mathcal{L} = \nabla_{\mathbf{o}_{residual}} \mathcal{L} \cdot (1 + \frac{\partial \mathcal{F}}{\partial \mathbf{x}})$. These blocks capture both micro-expressions ($\delta t < 500$ ms) and higher-level abstractions (gestures $\mathbf{G}(t) = \{\mathbf{p}_1(t),...,\mathbf{p}_J(t)\}$) with computational efficiency $\mathcal{O}(H'W'(k^2C' + C'^2))$.

\begin{algorithm}[t]
\caption{AVT-CA: Audio-Video Transformer Fusion with Cross Attention for MER}
\label{alg:AVTCA-alg}
\begin{algorithmic}[1]

\STATE \textbf{Input:} Audio $\mathbf{A} \in \mathbb{R}^{T_a \times C_a}$, Video $\mathbf{V} \in \mathbb{R}^{H \times W \times C_v}$
\vspace{5pt}

\STATE \textbf{Output:} Predicted emotion class $\hat{y}$
\vspace{5pt}

\STATE \textbf{1. Audio Feature Extraction:}
\STATE $\mathbf{\bar{A}} \gets \text{MaxPool}(\text{ReLU}(\text{BatchNorm}(\text{Conv1D}(\mathbf{A}, \mathbf{W}_a, \mathbf{b}_a))))$
\STATE Repeat above twice.
\vspace{5pt}

\STATE \textbf{2. Video Feature Extraction:}
\STATE $\mathbf{\bar{V}} \gets \text{MaxPool}(\text{ReLU}(\text{Conv2D}(\mathbf{V}, \mathbf{W}_v, \mathbf{b}_v))))$
\STATE $\mathbf{\tilde{V}}_c \gets \sigma(\text{FC}(\text{AvgPool}(\mathbf{\bar{V}})))$  \hfill \COMMENT{$attention_{channel}$}
\STATE $\mathbf{\tilde{V}}_s \gets \text{Softmax}(\text{Conv2D}(\mathbf{\bar{V}}, \mathbf{W}_s, \mathbf{b}_s)))$  \hfill \COMMENT{$attention_{spatial}$}
\STATE $\mathbf{\tilde{V}} \gets \mathbf{\tilde{V}}_c \odot \mathbf{\tilde{V}}_s$
\STATE $\mathbf{o}_{residual} \gets \text{DepthwiseConv}(\mathbf{\tilde{V}}) + \text{PointwiseConv}(\text{DepthwiseConv}(\mathbf{\tilde{V}}))$
\vspace{5pt}

\STATE \textbf{3. Transformer-based Cross-Modal Learning:}
\STATE $Q, K, V \gets XW_q, XW_k, XW_v$
\STATE $o_{attn} \gets \text{Softmax}(\frac{QK^\top}{\sqrt{d_k}}) V$
\STATE $o_{final} \gets o_{attn} + X + \text{FFN}(o_{attn} + X)$
\vspace{5pt}

\STATE \textbf{4. Cross-Self-Attention Fusion:}
\STATE $o_{AV} \gets \text{Softmax}(\frac{Q_a K_v^\top}{\sqrt{d_k}}) V_v$
\STATE $o_{VA} \gets \text{Softmax}(\frac{Q_v K_a^\top}{\sqrt{d_k}}) V_a$
\STATE $o'_{audio} \gets o_{AV} + o_{\text{self,audio}}$
\STATE $o'_{video} \gets o_{VA} + o_{\text{self,video}}$
\vspace{5pt}

\STATE \textbf{5. Emotion Prediction:}
\vspace{5pt}
\STATE $o^{\text{pool}}_{\text{audio}} \gets \max(o'_{\text{audio}})$
\vspace{5pt}
\STATE $o^{\text{pool}}_{\text{video}} \gets \max(o'_{\text{video}})$
\vspace{5pt}
\STATE $o^{\text{final}} \gets o^{\text{pool}}_{\text{audio}} + o^{\text{pool}}_{\text{video}}$
\STATE Compute $p(y | x) = \frac{\exp(z_i)}{\sum_j^C \exp(z_j)}$, where $z_i$ is the logit for class $i$
\vspace{5pt}

\STATE \textbf{Return} $\hat{y}$
\end{algorithmic}
\end{algorithm}

\subsection{Transformer Blocks for Cross-Modal Learning}

The refined audio features $\mathbf{\bar{A}}_2 \in \mathbb{R}^{T_a' \times C_a'}$ and video features $\mathbf{o}_{residual} \in \mathbb{R}^{T_v' \times C_v'}$ are passed into transformer blocks for cross-modal interaction $\mathcal{T}: \mathbb{R}^{T_a' \times C_a'} \times \mathbb{R}^{T_v' \times C_v'} \rightarrow \mathbb{R}^{T_a' \times d} \times \mathbb{R}^{T_v' \times d}$, enabling the model to capture relationships between audio and video cues through attention mapping $\mathcal{A}_{cross}: \mathcal{X}_A \times \mathcal{X}_V \rightarrow \mathcal{Y}_{AV}$.

\paragraph{Self-Attention Mechanism}
The self-attention mechanism dynamically weighs input sequence importance through similarity function $\mathcal{S}(x_i, x_j) = \frac{x_i \cdot x_j}{\|x_i\| \cdot \|x_j\|}$. For input $\mathbf{X} \in \mathbb{R}^{T \times d}$, self-attention computes:
\begin{equation}
\mathbf{o}_{attn} = \textbf{Softmax}\left(\frac{\mathbf{QK}^\top}{\sqrt{d_k}}\right)\mathbf{V} = \sum_{j=1}^{T} \frac{\exp(\frac{\mathbf{Q}_i \cdot \mathbf{K}_j^{\top}}{\sqrt{d_k}})}{\sum_{l=1}^{T} \exp(\frac{\mathbf{Q}_i \cdot \mathbf{K}_l^{\top}}{\sqrt{d_k}})} \mathbf{V}_j,
\end{equation}
where $\mathbf{Q} = \mathbf{X}\mathbf{W}_q$, $\mathbf{K} = \mathbf{X}\mathbf{W}_k$, and $\mathbf{V} = \mathbf{X}\mathbf{W}_v$ with learnable weights $\mathbf{W}_q, \mathbf{W}_k, \mathbf{W}_v \in \mathbb{R}^{d \times d_k}$. The scaling factor $d_k = d / h$ ensures numerical stability. This mechanism focuses on modality-specific emotional cues such as pitch changes indicative of anger $\mathcal{A}_{\text{anger}}(t) = \{t | |\nabla^2 p(t)| > \tau_a\}$ or facial micro-expressions $\mathcal{V}_{\text{emotion}}(t) = \{(i,j,t) | \mathcal{E}(i,j,t) > \tau_e\}$.

\paragraph{Multi-Head Attention}
Multi-head attention $\mathcal{M}: \mathbb{R}^{T \times d} \rightarrow \mathbb{R}^{T \times d}$ captures diverse subspace representations through $h$ parallel heads:
\begin{subequations}
    \begin{align}
    & \textbf{o}_i = \text{Attention}(\mathbf{X}\mathbf{W}_q^i, \mathbf{X}\mathbf{W}_k^i, \mathbf{X}\mathbf{W}_v^i), \quad i=1,\dots,h,\\
    & \textbf{o}_{multi} = \text{Concat}(\textbf{o}_1, \textbf{o}_2, \dots, \textbf{o}_h)\mathbf{W}_o,
    \end{align}
\end{subequations}
where $\mathbf{W}_o \in \mathbb{R}^{hd_k \times d}$ is a projection matrix. This allows simultaneous attention to multiple aspects through joint distribution $p(\mathbf{y} | \mathbf{X}) = \prod_{i=1}^{h} p_i(\mathbf{y}_i | \mathbf{X})$, enabling cross-modal synchronization via alignment function $\mathcal{T}_{align}(a_t, v_t) = \sum_{i=1}^{T} \alpha_i \cdot \delta(a_i, v_i)$.

\paragraph{Residual Connections and Normalization}
Residual connections prevent vanishing gradients: $\textbf{o}_{trans} = \textbf{o}_{attn} + \textbf{o}_{input}$. Layer normalization stabilizes learning:
\begin{equation}
\textbf{o}_{norm} = \gamma \odot \frac{\textbf{o}_{trans} - \mu(\textbf{o}_{trans})}{\sqrt{\sigma^2(\textbf{o}_{trans}) + \epsilon}} + \beta,
\end{equation}
where $\mu(\textbf{o}_{trans}) = \frac{1}{d}\sum_{j=1}^{d} \textbf{o}_{trans,j}$, $\sigma^2(\textbf{o}_{trans}) = \frac{1}{d}\sum_{j=1}^{d} (\textbf{o}_{trans,j} - \mu)^2$, and $\gamma, \beta \in \mathbb{R}^d$ are learnable parameters.

\paragraph{Feedforward Network (FFN)}
A feedforward network $\mathcal{F}_{FFN}: \mathbb{R}^{T \times d} \rightarrow \mathbb{R}^{T \times d}$ adds non-linearity:
\begin{subequations}
\begin{align}
& \textbf{o}_{ffn1} = \textbf{GELU}(\textbf{o}_{norm}\mathbf{W}_1 + \mathbf{b}_1),\\
& \textbf{o}_{ffn2} = \textbf{Dropout}(\textbf{o}_{ffn1}, p),\\
& \textbf{o}_{ffn3} = \textbf{o}_{ffn2}\mathbf{W}_2 + \mathbf{b}_2,
\end{align}
\end{subequations}
where $\mathbf{W}_1 \in \mathbb{R}^{d \times d_{ff}}$, $\mathbf{W}_2 \in \mathbb{R}^{d_{ff} \times d}$, $\mathbf{b}_1 \in \mathbb{R}^{d_{ff}}$, and $\mathbf{b}_2 \in \mathbb{R}^{d}$ with $d_{ff} = 4d$. GELU activation $\text{GELU}(x) = x \cdot \Phi(x)$ improves gradient flow. The final output includes another residual connection:
\begin{equation}
\textbf{o}_{final} = \textbf{o}_{norm} + \textbf{o}_{ffn3}.
\end{equation}

\paragraph{Cross-Modal Interaction in Transformer Blocks}
In cross-modal interactions, one modality serves as query while the other provides key-value pairs. For audio-to-video interaction, audio features act as queries ($\mathbf{{Q}}a = \mathbf{\bar{A}}_2 \mathbf{W}_q^a$), while video features serve as keys ($\mathbf{{K}}_v = \mathbf{o}_{residual} \mathbf{W}_k^v$) and values ($\mathbf{{V}}_v = \mathbf{o}_{residual} \mathbf{W}_v^v$). This alignment relates audio cues with visual expressions through probability $p(a_i \sim v_j) = \frac{\exp(\frac{\mathbf{{Q}}a_i \cdot \mathbf{{K}}_{v_j}^{\top}}{\sqrt{d_k}})}{\sum_{l=1}^{T_v} \exp(\frac{\mathbf{{Q}}a_i \cdot \mathbf{{K}}_{v_l}^{\top}}{\sqrt{d_k}})}$. 

For video-to-audio interaction, video features act as queries ($\mathbf{{Q}}_v = \mathbf{o}_{residual} \mathbf{W}_q^v$), with audio features as keys ($\mathbf{{K}}_a = \mathbf{\bar{A}}_2 \mathbf{W}_k^a$) and values ($\mathbf{{V}}_a = \mathbf{\bar{A}}_2 \mathbf{W}_v^a$), enabling contextualization of visual gestures with speech characteristics. Processing both interactions in parallel creates bidirectional mapping $\mathcal{M}_{A \leftrightarrow V} = \{\mathcal{M}_{A \rightarrow V}, \mathcal{M}_{V \rightarrow A}\}$.

\subsection{Cross-Self-Attention Mechanism}
To enable deeper interaction between modalities, the AVT-CA model employs a cross-self-attention mechanism $\mathcal{C}_{self}: \mathbb{R}^{T_a \times d_a} \times \mathbb{R}^{T_v \times d_v} \rightarrow \mathbb{R}^{T_a \times d} \times \mathbb{R}^{T_v \times d}$ positioned between transformer blocks and final cross-attention layers. This allows each modality to attend to both its own features and those of the other modality through joint attention space $\mathcal{J} = \mathcal{S}_A \cup \mathcal{S}_V \cup \mathcal{C}_{A,V}$. In the audio-to-video cross-attention, audio features act as queries with video features serving as keys and values, where $\mathbf{o}_{AV} = \textbf{Softmax}(\frac{\mathbf{Q}_a \mathbf{K}_v^\top}{\sqrt{d_k}}) \mathbf{V}_v = \sum_{j=1}^{T_v} \alpha_{i,j}^{AV} \mathbf{V}_{v,j}$ with $\alpha_{i,j}^{AV} = \frac{\exp(\mathbf{s}_{i,j}^{AV})}{\sum_{l=1}^{T_v} \exp(\mathbf{s}_{i,l}^{AV})}$ and $\mathbf{s}_{i,j}^{AV} = \frac{\mathbf{Q}_{a,i} \cdot \mathbf{K}_{v,j}^{\top}}{\sqrt{d_k}}$. Similarly, for video-to-audio cross-attention, $\mathbf{o}_{VA} = \textbf{Softmax}(\frac{\mathbf{Q}_v \mathbf{K}_a^\top}{\sqrt{d_k}}) \mathbf{V}_a = \sum_{j=1}^{T_a} \alpha_{i,j}^{VA} \mathbf{V}_{a,j}$ where $\alpha_{i,j}^{VA} = \frac{\exp(\mathbf{s}_{i,j}^{VA})}{\sum_{l=1}^{T_a} \exp(\mathbf{s}_{i,l}^{VA})}$ and $\mathbf{s}_{i,j}^{VA} = \frac{\mathbf{Q}_{v,i} \cdot \mathbf{K}_{a,j}^{\top}}{\sqrt{d_k}}$.

These mechanisms align complementary information between modalities through attention mapping $\mathcal{M}_{align}(a,v) = \sum_{i=1}^{T_a}\sum_{j=1}^{T_v} \alpha_{i,j} \cdot \mathcal{S}(a_i, v_j)$, with correlation coefficient $\rho(a,v) = \frac{\text{Cov}(a,v)}{\sigma_a \sigma_v}$. Outputs are combined with inputs via residual connections as $\tilde{\mathbf{o}}_{audio} = \mathbf{o}_{AV} + \mathbf{o}_{self}^{res,audio} = \mathbf{o}_{self}^{res,audio} + \mathcal{F}_{AV}(\mathbf{o}_{self}^{audio}, \mathbf{o}_{self}^{video})$ and $\tilde{\mathbf{o}}_{video} = \mathbf{o}_{VA} + \mathbf{o}_{self}^{res,video} = \mathbf{o}_{self}^{res,video} + \mathcal{F}_{VA}(\mathbf{o}_{self}^{video}, \mathbf{o}_{self}^{audio})$, where $\mathcal{F}_{AV}$ and $\mathcal{F}_{VA}$ represent the cross-attention transformation functions with parameters $\{\mathbf{W}_q^a, \mathbf{W}_k^v, \mathbf{W}_v^v\}$ and $\{\mathbf{W}_q^v, \mathbf{W}_k^a, \mathbf{W}_v^a\}$ respectively.

\subsection{Final Emotion Prediction}

Finally, refined audio and video features are combined to produce the final emotion prediction. The outputs from audio $\tilde{\mathbf{o}}_{audio} \in \mathbb{R}^{T_a' \times d}$ and video $\tilde{\mathbf{o}}_{video} \in \mathbb{R}^{T_v' \times d}$ streams pass through cross-attention blocks with $h$ heads. Each head processes a portion of the feature space with dimension $d_k = \frac{d}{h}$ where $q_i = \mathbf{X}\mathbf{W}_q^i \in \mathbb{R}^{T \times d_k}$, $k_i = \mathbf{Y}\mathbf{W}_k^i \in \mathbb{R}^{T' \times d_k}$, and $v_i = \mathbf{Y}\mathbf{W}_v^i \in \mathbb{R}^{T' \times d_k}$ for $i \in \{1,2,...,h\}$. The attention output for each head is $o_i = \sum_{j=1}^{T'} \alpha_{i,j} v_j$ where $\alpha_{i,j} = \frac{\exp(\frac{q_i \cdot k_j^{\top}}{\sqrt{d_k}})}{\sum_{l=1}^{T'} \exp(\frac{q_i \cdot k_l^{\top}}{\sqrt{d_k}})}$. Outputs from all heads are concatenated as $o_{modality} = \text{Concat}(o_1, o_2, ..., o_h)\mathbf{W}_o \in \mathbb{R}^{T \times d}$, where $\mathbf{W}_o \in \mathbb{R}^{hd_k \times d}$ is a projection matrix. This configuration captures various emotional cues through parallel attention paths $\{\mathcal{A}_1, \mathcal{A}_2, ..., \mathcal{A}_h\}$.

Max pooling reduces dimensionality while retaining salient features, with $\mathbf{o}_{audio}^{pool} = \max_{i \in \{1,2,...,T_a'\}} (\mathbf{o}_{audio}[i]) \in \mathbb{R}^d$ and $\mathbf{o}_{video}^{pool} = \max_{j \in \{1,2,...,T_v'\}} (\mathbf{o}_{video}[j]) \in \mathbb{R}^d$. The pooled representations are combined via element-wise addition as $\mathbf{o}_{final} = \mathbf{o}_{audio}^{pool} + \mathbf{o}_{video}^{pool} \in \mathbb{R}^d$. This combined representation passes through a fully connected layer and softmax activation where $z = \mathbf{W}_{fc} \cdot \mathbf{o}_{final} + \mathbf{b}_{fc} \in \mathbb{R}^C$ and $p(y|\mathbf{A}, \mathbf{V}) = \{\frac{\exp(z_c)}{\sum_{j=1}^{C}\exp(z_j)} | c \in \{1,2,...,C\}\} \in \mathbb{R}^C$. The final predicted emotion is determined by selecting the class with highest probability as $\hat{y} = \arg\max_{c \in \{1,2,...,C\}} p(y_c|\mathbf{A}, \mathbf{V})$.

The model is trained using cross-entropy loss $\mathcal{L}_{CE}(\mathbf{Y}, \hat{\mathbf{Y}}) = -\frac{1}{N} \sum_{i=1}^{N} \sum_{c=1}^{C} y_{i,c} \log(\hat{y}_{i,c})$, where $C$ is the number of emotion categories, $N$ is the batch size, $y_{i,c}$ is a binary indicator (1 if instance $i$ belongs to class $c$, 0 otherwise), and $\hat{y}_{i,c}$ is the predicted probability. Model parameters $\theta$ are optimized via stochastic gradient descent with adaptive learning rate through update rule $\theta_{t+1} = \theta_t - \eta(t) \nabla_\theta \mathcal{L}_{CE}(\mathbf{Y}, \hat{\mathbf{Y}}; \theta_t)$.

\section{Experimental Setup}

The AVT-CA model and baseline models were trained for 128 epochs using a high-performance system with the following specifications: the operating system was Linux 5.15.133, and the CPU used was an AMD EPYC 7763 64-core processor with a processor architecture of $X_{86_{64}}$. The system had 128 CPU(s) with a CPU family of 25, model 1, one thread per core, 64 cores per socket, and two sockets. The hyperparameters used for training all models were consistent across experiments to ensure a fair comparison. The batch size was set to 8, and the Adam optimizer was used with a learning rate of 0.01 and a weight decay of 0.001. AVT-CA underwent an extensive training process lasting approximately 72 hours to ensure convergence and optimal performance. To validate the effectiveness of the proposed AVT-CA model, all baseline models described in Sections~4.3, 4.4, and 4.5 were trained under identical experimental conditions. This cross-validation approach ensures that the comparative evaluation is fair and unbiased.

\subsection{Dataset Details}

The experiments utilize three standard multimodal emotion recognition datasets:

\subsubsection{RAVDESS \cite{51}} 
The Ryerson Audio-Visual Database contains 7,356 files (24.8 GB) from 24 professional actors (12 female, 12 male) expressing eight emotions (calm, happy, sad, angry, fearful, surprise, disgust, and neutral) at two intensity levels. Files are available in audio-only, audio-video, and video-only formats, making it a valuable benchmark for emotion analysis.

\begin{figure}[t]
\centering
\includegraphics[width=\columnwidth]{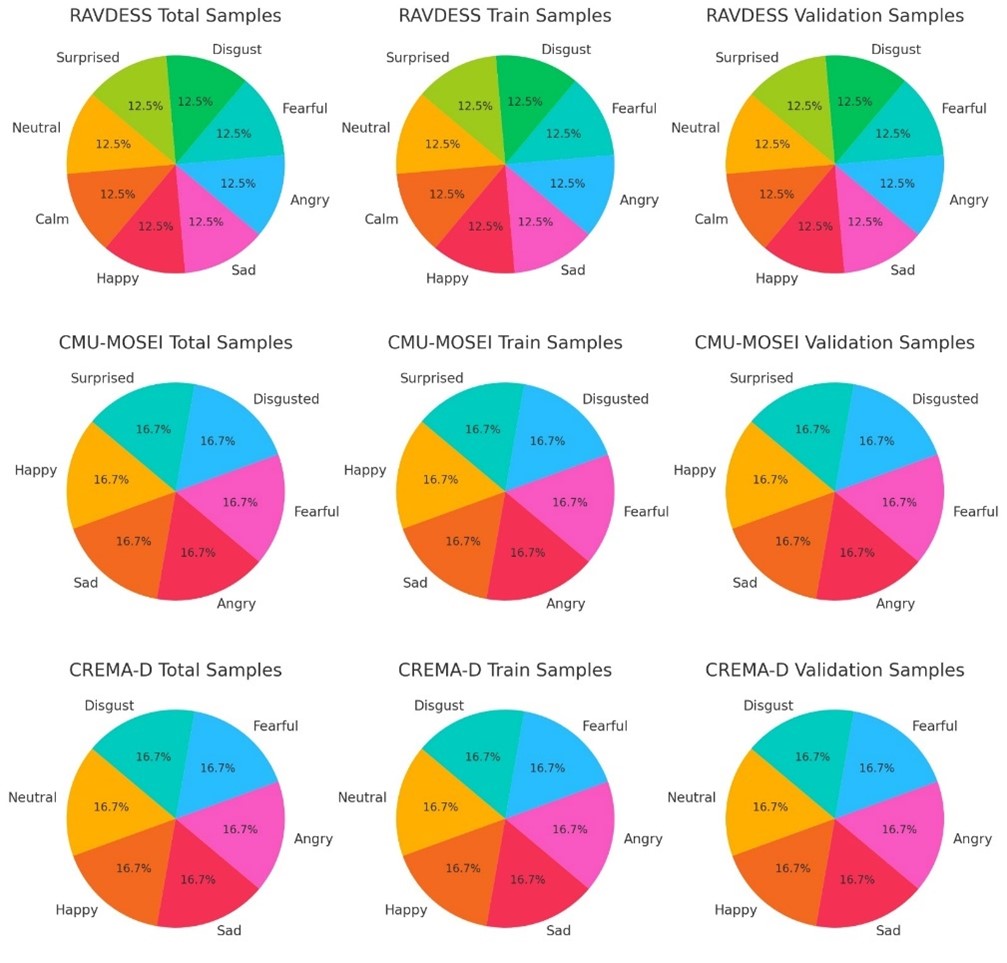}
\caption{
Distribution of emotion categories across datasets and splits.
Each row shows one benchmark (RAVDESS, CMU-MOSEI, CREMA-D), while columns report
the label distribution for the full dataset, training set, and validation set.
All datasets are approximately balanced across classes, indicating that models trained on these data
are not biased toward a particular emotion.
}
\label{fig:datasetStatistics}
\end{figure}

\subsubsection{CMU-MOSEI \cite{52}} 
This dataset comprises over 23,500 sentence utterance videos from 1,000+ diverse speakers on YouTube, with 65+ hours of annotated content. It features punctuated transcriptions and balanced gender representation across six emotions (happy, sad, angry, fearful, disgusted, surprised), serving as a comprehensive resource for multimodal sentiment and emotion analysis.

\subsubsection{CREMA-D \cite{53}} 
The Crowd-sourced Emotional Multimodal Actors Dataset includes 7,442 clips from 91 actors (48 male, 43 female, aged 20-74) from diverse racial backgrounds. Actors perform 12 sentences with six emotions (anger, disgust, fear, happy, neutral, sad) at four intensity levels, with ratings based on audiovisual, video-only, and audio-only presentations.

\begin{table*}[h]
\centering
\caption{Dataset Statistics for Multimodal Emotion Recognition}
\resizebox{\textwidth}{!}{%
\begin{tabular}{l|ccc|ccc|ccc}
\toprule
& \multicolumn{3}{c|}{\textbf{RAVDESS}} & \multicolumn{3}{c|}{\textbf{CMU-MOSEI}} & \multicolumn{3}{c}{\textbf{CREMA-D}} \\
\textbf{Emotion} & \textbf{Total} & \textbf{Train (80\%)} & \textbf{Val (20\%)} & \textbf{Total} & \textbf{Train (80\%)} & \textbf{Val (20\%)} & \textbf{Total} & \textbf{Train (80\%)} & \textbf{Val (20\%)} \\
\midrule
Neutral   & 920  & 736  & 184 & - & - & - & 1240 & 992 & 248 \\
Calm      & 920  & 736  & 184 & - & - & - & - & - & - \\
Happy     & 920  & 736  & 184 & 3909 & 3127 & 782 & 1240 & 992 & 248 \\
Sad       & 920  & 736  & 184 & 3909 & 3127 & 782 & 1240 & 992 & 248 \\
Angry     & 920  & 736  & 184 & 3909 & 3127 & 782 & 1240 & 992 & 248 \\
Fearful   & 920  & 736  & 184 & 3909 & 3127 & 782 & 1240 & 992 & 248 \\
Disgust   & 920  & 736  & 184 & 3909 & 3127 & 782 & 1240 & 992 & 248 \\
Surprised & 920  & 736  & 184 & 3909 & 3127 & 782 & - & - & - \\
\midrule
\textbf{Total} & \textbf{7360} & \textbf{5888} & \textbf{1472} & \textbf{23454} & \textbf{18763} & \textbf{4691} & \textbf{7440} & \textbf{5952} & \textbf{1488} \\
\bottomrule
\end{tabular}%
}
\label{tab:datasets}
\end{table*}

The data distribution across each split and dataset is shown in Figure~\ref{fig:datasetStatistics}.

\subsection{Evaluation Metrics}
Evaluation metrics like classification accuracy and F1-score are critical for assessing deep learning models in emotion recognition. Accuracy measures the model's ability to precisely identify emotional states from various data modalities (audio, video). The F1-score evaluates the balance between precision (correct positive predictions among all predicted positives) and recall (correct positive predictions among all actual positives), offering a consolidated measure of classification performance. High F1 scores indicate effective emotion classification, considering both false positives and false negatives, crucial for reliable emotion recognition systems.

\subsection{Baseline Models for RAVDESS dataset}

\subsubsection{Vector Quantized Masked Autoencoder (VQ-MAE) \cite{43}}  
VQ-MAE is a model designed for MER that merges vector quantization with masked autoencoding techniques. It learns discrete latent representations from multimodal inputs such as audio, video, and text by masking sections of the input data and reconstructing them. This process enhances the robustness and informativeness of the extracted features. The vector quantization discretizes continuous inputs, effectively capturing diverse emotional cues across modalities. Utilizing self-supervised learning, VQ-MAE improves performance in emotion recognition tasks while minimizing the need for labeled data.

\subsubsection{Multimodal Transfer Module (MMTM) \cite{60}}  
MMTM is a framework for MER that facilitates information exchange between different modalities, including audio, video, and text. It dynamically refines the feature representations of each modality through cross-modal attention mechanisms, enabling the model to effectively leverage complementary information from different sources.

\subsubsection{Multimodal Self-Attention Fusion (MSAF) \cite{36}}  
MSAF is a technique for MER that incorporates self-attention mechanisms across multiple modalities such as audio, video, and text. By dynamically weighting and fusing features from each modality based on their relevance to the emotional context, MSAF enhances the model's ability to capture nuanced emotional expressions.

\begin{figure*}[t]
    \centering
    \subfloat[Accuracy curve of AVT-CA on the RAVDESS dataset\label{img:ravAcc}]{
        \includegraphics[width=0.3\textwidth]{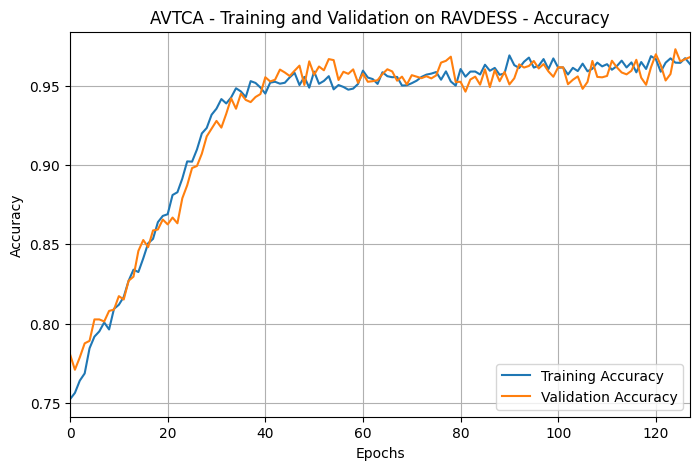}
    }\hfill
    \subfloat[F1-Score curve of AVT-CA on the RAVDESS dataset\label{img:ravF1}]{
        \includegraphics[width=0.3\textwidth]{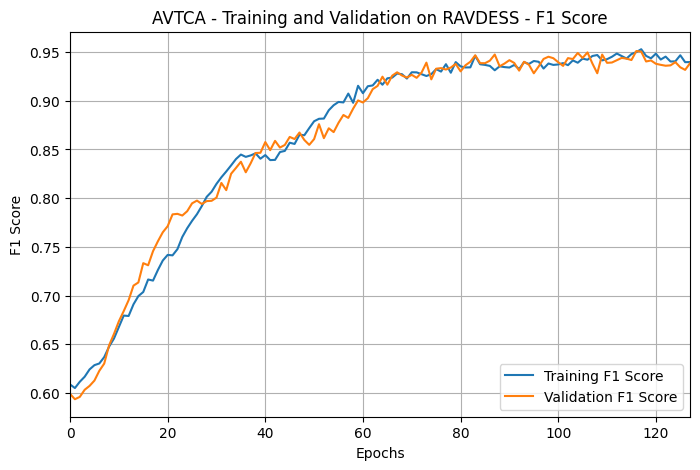}
    }\hfill
    \subfloat[Loss curve of AVT-CA on the RAVDESS dataset\label{img:ravLoss}]{
        \includegraphics[width=0.3\textwidth]{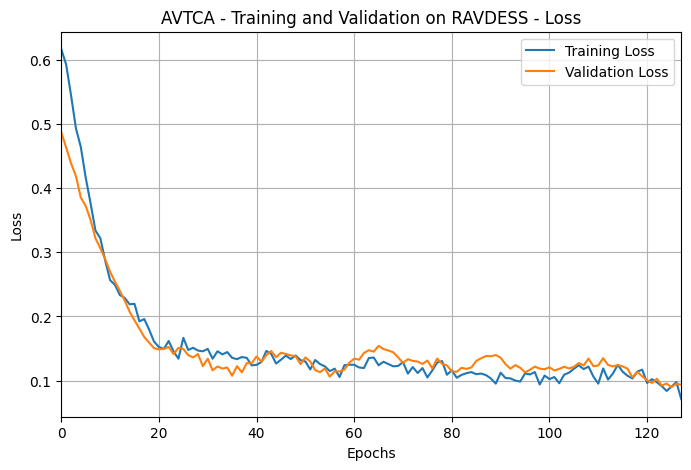}
    }
    \caption{Performance curves of AVT-CA on the RAVDESS dataset, illustrating accuracy, F1-Score, and loss behavior during training and validation.}
    \label{fig:ravResults}
\end{figure*}

\begin{figure*}[t]
    \centering
    \subfloat[Accuracy curve of AVT-CA on the CMU-MOSEI dataset\label{img:cmuAcc}]{
        \includegraphics[width=0.3\textwidth]{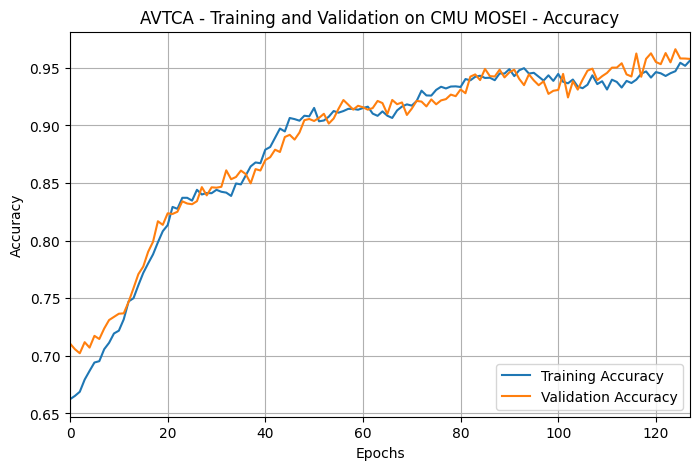}
    }\hfill
    \subfloat[F1-Score curve of AVT-CA on the CMU-MOSEI dataset\label{img:cmuF1}]{
        \includegraphics[width=0.3\textwidth]{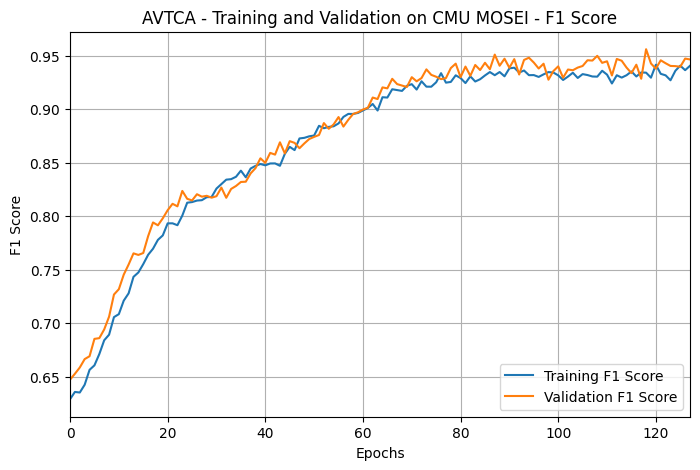}
    }\hfill
    \subfloat[Loss curve of AVT-CA on the CMU-MOSEI dataset\label{img:cmuLoss}]{
        \includegraphics[width=0.3\textwidth]{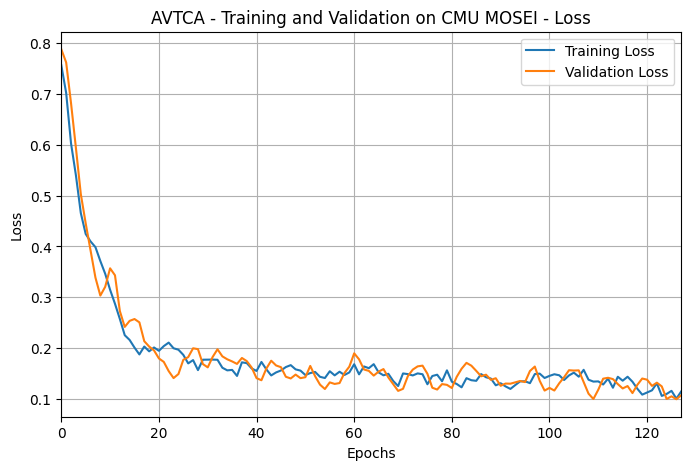}
    }
    \caption{Performance curves of AVT-CA on the CMU-MOSEI dataset, illustrating accuracy, F1-Score, and loss behavior during training and validation.}
    \label{fig:cmuResults}
\end{figure*}

\subsubsection{Cross-Modal Fusion Network with Self-Regularization (CFN-SR) \cite{50}}  
CFN-SR is a model designed for MER that leverages cross-modal fusion techniques alongside self-regularization mechanisms. It integrates features from different modalities, such as audio, video, and text, using neural networks to enhance the synergy between modalities. Additionally, CFN-SR incorporates self-regularization, improving generalization capability and resilience against noisy or incomplete data, thereby strengthening its robustness in emotion recognition tasks.

\subsubsection{Convolutional Neural Network (CNN) \cite{61}}  
This approach investigates emotion recognition using a multimodal framework based on convolutional neural networks (CNNs). It combines facial frames, optical flow, and Mel spectrograms extracted from videos to identify the most effective modality combination for recognizing emotions.

\subsubsection{Multimodal Masked Autoencoder for Dynamic Emotion Recognition (MultiMAE-DER) \cite{62}}  
MultiMAE-DER is a model that leverages spatiotemporal correlations across visual and audio modalities for dynamic emotion recognition. Utilizing a pre-trained masked autoencoder, it is fine-tuned to optimize performance. MultiMAE-DER employs six fusion strategies to process multimodal input sequences, effectively capturing spatial, temporal, and spatiotemporal feature correlations.

\subsection{Baseline Models for CMU-MOSEI dataset}
\subsubsection{Low-rank Multimodal Fusion (LMF) \cite{47}} 
LMF is a technique for emotion recognition that combines audio, video, and text features by constructing a high-dimensional tensor and applying low-rank tensor factorization. This method efficiently reduces the tensor's dimensionality while capturing critical inter-modality interactions with reduced computational complexity. 

\subsubsection{Memory Fusion Network (MFN) \cite{54}} 
MFN is designed for MER and utilizes a memory matrix to encode long-term inter-modality interactions across sequences. It processes each modality independently and employs a multi-view gated memory mechanism to integrate and retain these interactions effectively.

\subsubsection{Multimodal Factorization Model (MFM) \cite{55}}  
MFM decomposes audio, video, and text input features into shared and modality-specific components using tensor factorization. This approach effectively captures both common characteristics and unique aspects of each modality, facilitating robust fusion for emotion recognition.

\subsubsection{Multimodal Transformer (MulT) \cite{56}}  
MulT is a transformer-based model for emotion recognition that processes temporal sequences of multimodal data, including audio, video, and text. It incorporates cross-modal attention mechanisms to capture interactions between modalities by aligning and integrating their respective features. This approach enhances contextual understanding as each modality attends to and benefits from the others' information.

\subsubsection{MAG-BERT \cite{57}}  
MAG-BERT extends BERT for emotion recognition by integrating multimodal features such as audio and video into its textual embeddings. It utilizes adaptation gates to selectively incorporate non-verbal information into BERT’s layers, maintaining the multimodal context while filtering out noise. This ensures effective fusion of modalities without compromising BERT’s pre-trained language representations.

\subsubsection{ConCluGen \cite{58}}  
ConCluGen applies multitask multimodal self-supervised learning for facial expression recognition in in-the-wild video data. It integrates three self-supervised objectives: multimodal contrastive loss, multimodal clustering loss, and multimodal data reconstruction loss, ensuring a comprehensive representation of facial expressions.

\subsubsection{Contextualized Graph Neural Network-based MER (COGMEN) \cite{59}}  
COGMEN is a system designed for MER that leverages both local (inter/intra-speaker dependencies) and global (context) information. It employs a Graph Neural Network (GNN) architecture to model these complex dependencies within conversations, improving the contextual understanding of emotional expressions.

\begin{figure*}[ht]
    \centering
    \subfloat[Accuracy curve of AVT-CA on the CREMA-D dataset\label{img:cremaAcc}]{
        \includegraphics[width=0.3\textwidth]{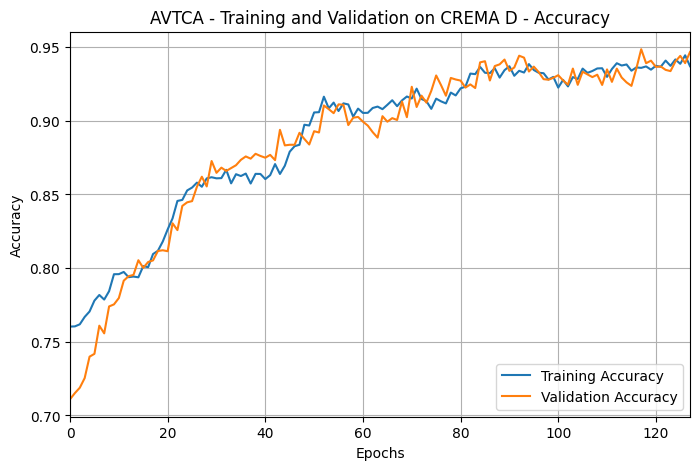}
    }\hfill
    \subfloat[F1-Score curve of AVT-CA on the CREMA-D dataset\label{img:cremaF1}]{
        \includegraphics[width=0.3\textwidth]{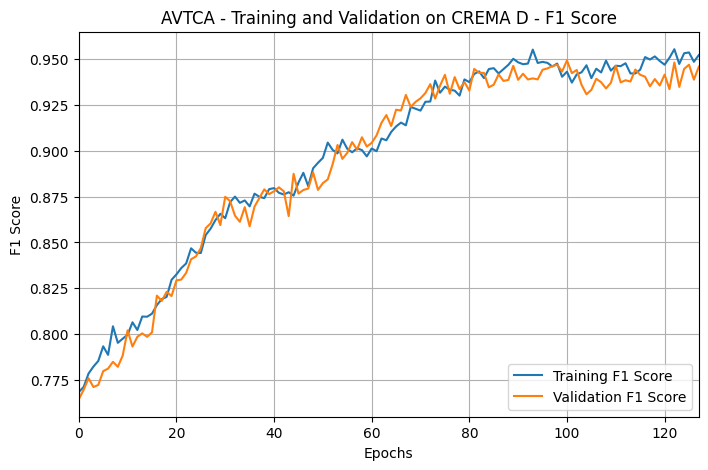}
    }\hfill
    \subfloat[Loss curve of AVT-CA on the CREMA-D dataset\label{img:cremaLoss}]{
        \includegraphics[width=0.3\textwidth]{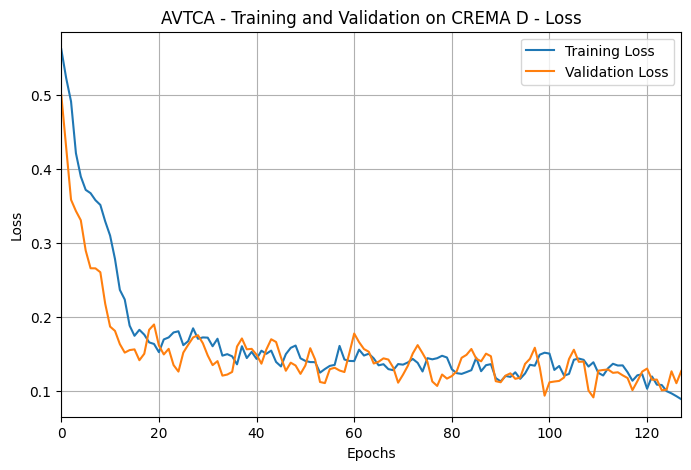}
    }
    \caption{Performance curves of AVT-CA on the CREMA-D dataset, illustrating accuracy, F1-Score, and loss behavior during training and validation.}
    \label{fig:cremaResults}
\end{figure*}

\subsection{Baseline Models for CREMA-D dataset}
\subsubsection{AuxFormer \cite{63}}  
AuxFormer is a framework developed for MER that merges transformer architectures with auxiliary tasks to exploit additional modalities. It employs transformer encoders to individually process textual, audio, and visual inputs. Auxiliary tasks are integrated to encourage the model to learn representations that are advantageous for emotion recognition.

\subsubsection{Versatile-Audio-Visual Learning (VAVL) \cite{64}}  
VAVL is a model tailored for MER that incorporates visual, audio, and verbal inputs through Long Short-Term Memory (LSTM) networks. It captures temporal dependencies and interactions across these modalities to accurately predict emotional states. VAVL employs hierarchical fusion mechanisms to integrate multimodal features at various levels, thereby improving its capacity to interpret intricate emotional expressions.

\subsubsection{Latent Attention-based Dual Decoder Representation (LADDER) \cite{65}}  
LADDER is a model designed for MER that incorporates dual decoders along with attention mechanisms. It extracts latent representations from each modality - audio, video, and text - and utilizes attention to effectively integrate these features. LADDER combines hierarchical fusion with latent representations to capture nuanced emotional cues across different modalities.

\subsubsection{Dynamic Emotion Integration with Information Interaction (DE III) \cite{66}}  
DE III is a framework developed for MER that dynamically integrates emotional information across modalities such as audio, video, and text. It employs information interaction mechanisms to enhance the fusion of multimodal features, with a focus on capturing temporal dependencies and contextual clues. DE III utilizes deep learning architectures to optimize the integration process, thereby improving the model’s capability to interpret and classify complex emotional states effectively.

\subsubsection{Multimodal Transformer \cite{67}}  
The Multimodal Transformer model employs three transformer branches to compute audio self-attention, video self-attention, and audio-video cross-attention, identifying the most relevant information in each modality. This approach demonstrated superior performance in ablation studies. Additionally, a novel temporal embedding scheme, termed block embedding, is introduced to incorporate temporal information into visual features derived from multiple video frames.

\section{Results and Discussion}

This section evaluates AVT-CA's performance against baseline models across three standard datasets. Tables~\ref{tab:ravdess-results}, \ref{tab:cmu-mosei-results}, and \ref{tab:crema-d-results} present comprehensive comparison results.

\begin{table*}[t]
\centering
\begin{tabular}{@{}c@{}}
\begin{minipage}{0.32\textwidth}
\centering
\caption{Results on RAVDESS}
\begin{tabular}{lcc}
\toprule
\textbf{Model} & \textbf{Acc (\%)} & \textbf{F1 (\%)} \\
\midrule
MultiMAE-DER   & 69.17 & 69.03 \\
MMTM           & 73.12 & 72.70 \\
MSAF           & 74.86 & 74.31 \\
CFN-SR         & 75.76 & 75.27 \\
VQ-MAE         & 84.10 & 84.40 \\
CNN            & 95.95 & 92.17 \\
\textbf{AVT-CA} & \textbf{96.11} & \textbf{93.78} \\
\bottomrule
\end{tabular}
\label{tab:ravdess-results}
\end{minipage}
\hfill
\begin{minipage}{0.32\textwidth}
\centering
\caption{Results on CMU-MOSEI}
\begin{tabular}{lcc}
\toprule
\textbf{Model} & \textbf{Acc (\%)} & \textbf{F1 (\%)} \\
\midrule
ConCluGen      & 66.01 & 66.30 \\
MFN            & 76.03 & 76.07 \\
LMF            & 82.01 & 82.13 \\
MuIT           & 82.51 & 82.31 \\
COGMEN         & 83.81 & 83.51 \\
MFM            & 84.41 & 84.36 \\
MAG-BERT       & 84.71 & 84.51 \\
\textbf{AVT-CA} & \textbf{95.84} & \textbf{94.13} \\
\bottomrule
\end{tabular}
\label{tab:cmu-mosei-results}
\end{minipage}
\hfill
\begin{minipage}{0.32\textwidth}
\centering
\caption{Results on CREMA-D}
\begin{tabular}{lcc}
\toprule
\textbf{Model} & \textbf{Acc (\%)} & \textbf{F1 (\%)} \\
\midrule
Multi. Trans.  & 72.60 & 71.40 \\
AuxFormer      & 76.30 & 69.80 \\
VAVL           & 82.60 & 77.90 \\
LADDER         & 80.30 & 80.20 \\
DE III         & 83.70 & 79.50 \\
\textbf{AVT-CA} & \textbf{94.13} & \textbf{94.67} \\
\bottomrule
\end{tabular}
\label{tab:crema-d-results}
\end{minipage}
\end{tabular}
\end{table*}

\subsection{Result Analysis on RAVDESS dataset}

On the RAVDESS dataset, MultiMAE-DER \cite{62} (69.17\% accuracy) failed to capture key spatiotemporal correlations. MMTM \cite{60} (73.12\%), MSAF \cite{36} (74.86\%), and CFN-SR \cite{50} (75.76\%) showed moderate performance due to inefficient modality information transfer and feature misinterpretation. VQ-MAE \cite{43} achieved better results (84.10\%), while the CNN-based approach \cite{61} performed strongly (95.95\%) by effectively combining facial frames, optical flow, and Mel spectrograms. AVT-CA outperformed all baselines (96.11\% accuracy, 93.78\% F1-score) through its cross-attention mechanism that captures key audio-video interactions, adaptive feature refinement to preserve crucial emotional cues, and effective spatiotemporal alignment for contextual dependency preservation.

\subsection{Result Analysis on CMU-MOSEI dataset}

On CMU-MOSEI, ConCluGen \cite{58} performed poorly (66.01\% accuracy) due to limitations in capturing multimodal interactions. MFN \cite{54} achieved 76.03\% accuracy but showed inadequate fusion of spatial-temporal information. LMF \cite{47} (82.01\%), MuIT \cite{56} (82.51\%), and COGMEN \cite{59} (83.81\%) demonstrated moderate results but suffered from oversimplified modal interactions, sensitivity to noisy modalities, and limited video feature capture respectively. MFM \cite{55} (84.41\%) and MAG-BERT \cite{57} (84.71\%) showed competitive performance through effective factorized bilinear pooling and conditional BERT representations. AVT-CA significantly outperformed all baselines (95.84\% accuracy, 94.13\% F1-score) through its advanced feature representation techniques, fine-grained multimodal interactions without oversimplification, adaptive modality weighting to mitigate noise impact, and hierarchical graph-based processing for capturing subtle feature relationships.

\subsection{Result Analysis on CREMA-D Dataset}

On CREMA-D, Multimodal Transformer \cite{67} achieved 72.6\% accuracy but failed to extract diverse features across modalities. AuxFormer \cite{63} showed moderate performance (76.3\%) but was misled by auxiliary data during fusion. VAVL \cite{64}, LADDER \cite{65}, and DE III \cite{66} performed better (>80\% accuracy) by integrating contextual emotional cues, but their broader context limited focus on critical modality-specific features. AVT-CA substantially outperformed all baselines (94.13\% accuracy, 94.67\% F1-score) through efficient feature representation, cross-attention-based fusion for rich inter-modal dependencies, context-aware integration for robust multimodal alignment, and balanced preservation of both contextual and modality-specific information for comprehensive emotion representation.

\subsection{Ablation Study}
To analyze each module's contribution within our AVT-CA model, we evaluated five configurations: (i) Intermediate Transformer with single attention head (IT-1), (ii) Intermediate Transformer with four attention heads (IT-4), (iii) Cross Attention with single head (CT-1), (iv) Cross Attention with four heads (CT-4), and (v) the complete AVT-CA model integrating both IT-4 and CT-4, across all three datasets.

\begin{table}[t]
\centering
\caption{Ablation Study Results Across Datasets}
\resizebox{\columnwidth}{!}{%
\begin{tabular}{l|cc|cc|cc}
\toprule
& \multicolumn{2}{c|}{\textbf{RAVDESS}} & \multicolumn{2}{c|}{\textbf{CMU-MOSEI}} & \multicolumn{2}{c}{\textbf{CREMA-D}} \\
\textbf{Model} & \textbf{Acc (\%)} & \textbf{F1 (\%)} & \textbf{Acc (\%)} & \textbf{F1 (\%)} & \textbf{Acc (\%)} & \textbf{F1 (\%)} \\
\midrule
IT-1           & 76.41 & 76.31 & 67.72 & 67.13 & 71.50 & 70.90 \\
IT-4           & 78.50 & 78.31 & 64.91 & 63.97 & 74.80 & 74.30 \\
CT-1           & 76.00 & 75.90 & 64.94 & 64.33 & 72.10 & 71.40 \\
CT-4           & 77.41 & 77.13 & 67.72 & 67.91 & 75.50 & 74.90 \\
AVT-CA (Proposed) & \textbf{96.11} & \textbf{93.78} & \textbf{95.84} & \textbf{94.13} & \textbf{94.13} & \textbf{94.67} \\
\bottomrule
\end{tabular}%
}
\label{tab:ablation}
\end{table}

Experimental results show that IT-4 outperforms IT-1 on RAVDESS and CREMA-D, demonstrating the benefit of multiple attention heads in transformer fusion. These heads simultaneously focus on different input sequence aspects, capturing diverse emotional cues and patterns. Similarly, CT-4 enhances performance over CT-1 across all datasets, confirming the advantage of multiple attention heads in cross-attention for extracting significant features from both modalities. The complete AVT-CA model achieves state-of-the-art results across all datasets through synergistic integration: IT-4 captures interlinked audio-video features while CT-4 isolates and emphasizes critical features for emotion recognition. This significant performance improvement (approximately 20\% increase in accuracy and F1-score) demonstrates that each component is essential to the architecture's effectiveness in multimodal emotion recognition.

\subsection{Key Features of the Proposed AVT-CA Model}
The key features and highlights of the proposed model are as follows:
\begin{itemize}
    \item The cross-attention mechanism in the proposed AVT-CA model effectively learns salient information across modalities, which helps mitigate synchronization issues arising from temporal inconsistencies and disparate sampling rates.
    \item By leveraging spatial attention, channel attention, and a local feature extractor, the AVT-CA model's hierarchical feature refinement enhances its ability to extract both high-level and fine-grained emotional expressions.
    \item Through its cross-modality Transformer fusion, AVT-CA efficiently integrates enriched video features with audio data, enabling a structured and context-aware integration of multimodal information.
\end{itemize}

\section{Conclusion}
This paper presents AVT-CA, a novel approach to MER that leverages advanced feature representation and fusion techniques. The proposed model employs a unique feature extraction strategy incorporating channel attention, spatial attention, and a local feature extractor, which together enhance the critical features within input videos. This refined representation significantly improves the model’s ability to capture and emphasize subtle emotional cues, leading to more accurate emotion recognition. Additionally, AVT-CA introduces a transformer fusion mechanism designed to identify and align interlinked features between audio and video modalities. This mechanism effectively addresses synchronization issues, ensuring precise temporal alignment between audio and video inputs. To further enhance the model’s ability to discern complex emotional states, a cross-attention mechanism is implemented, selectively extracting and emphasizing key information from both modalities.

Empirical results demonstrate that AVT-CA achieves state-of-the-art performance in emotion recognition tasks, surpassing existing methods. An ablation study is performed to validate the significance of each module, indicating their individual contributions to the model’s overall effectiveness. Future work could integrate continual learning techniques to enable AVT-CA to adapt over time without catastrophic forgetting, making it more effective for real-world, ever-changing emotional contexts. Further research can also explore self-supervised or weakly supervised learning methods to allow AVT-CA to leverage large-scale unlabeled multimodal data, reducing dependence on manually labeled datasets.

\bibliographystyle{IEEEtran}
\bibliography{references} 

\end{document}